\pdfoutput=1
\documentclass[usenatbib,usegraphicx,useAMS,amsmath,amssymb]{mn2e}
\newcommand{\psiarc}{\psi_{\rm arc}}
\newcommand{\icell}{I_{\rm cell}}
\newcommand{\ipolar}{I_{\rm polar}}
\newcommand{\iarc}{I_{\rm arc}}
\newcommand{\psiplum}{\psi_{\rm Plum.}}
\newcommand{\psicell}{\psi_{\rm cell}}

\newcommand{\ain}{a_{\rm in}}

\newcommand{\mnras}{MNRAS}
\newcommand{\apj}{Ap. J.}
\newcommand{\apjs}{Ap. J. Supp. Ser.}
\newcommand{\aap}{Aston. \& Astrophys.}

\newcommand{\sinc}[1]{\rm sinc #1}
\renewcommand{\vec}[1]{\bmath #1}

\title[]{A truly Newtonian softening length for disc simulations}
\author[Jean-Marc Hur\'e and Audrey Trova]{J.-M. Hur\'e$^{1,2}$\thanks{E-mail:jean-marc.hure@obs.u-bordeaux1.fr} and A. Trova$^{2}$\thanks{E-mail:audrey.trova@asu.cas.cz}\\
$^{1}$Univ. Bordeaux, LAB, UMR 5804, F-33270, Floirac, France\\
$^{2}$CNRS, LAB, UMR 5804, F-33270, Floirac, France\\
$^{3}$Astronomical Institute of the Academy of Sciences, Bocni II 1401, 14131 Prague, Czech Republic}
\begin{document}

\date{Accepted 2014 November}

\pagerange{\pageref{firstpage}--\pageref{lastpage}} \pubyear{2002}

\maketitle

\label{firstpage}

\begin{abstract}
The softened point mass model is commonly used in simulations of gaseous discs including self-gravity while the value of associated length $\lambda$ remains, to some degree, controversial. This ``parameter'' is however fully constrained when, in a discretized disc, all fluid cells are demanded to obey Newton's law. We examine the topology of solutions in this context, focusing on cylindrical cells more or less vertically elongated. We find that not only the nominal length depends critically on the cell's shape (curvature, radial extension, height), but it is either a real or an imaginary number. Setting $\lambda$ as a fraction of the local disc thickness | as usually done | is indeed not the optimal choice. We then propose a novel prescription valid irrespective of the disc properties and grid spacings. The benefit, which amounts to $2-3$ more digits typically, is illustrated in a few concrete cases. A detailed mathematical analysis is in progress.
\end{abstract}

\begin{keywords}
accretion, accretion discs -- gravitation -- methods: numerical

\end{keywords}

\section{Motivations and objectives}

In simulations where a fluid is discretized on a computational grid, one must preliminary decide the way to treat matter contained inside each ``numerical cell'', and how the physical quantities are assigned to the nodes. The problem is tricky when considering self-gravity because the Newtonian potential, namely at point P$(\vec{r})$ \citep[e.g.][]{kellogg29}
\begin{equation}
\psicell(\vec{r})= -G\int_{\rm cell}{\frac{dm'(\vec{r}')}{|\vec{r}-\vec{r}'|}},
\label{eq:psicell}
\end{equation}
where $dm'$ is the elementary mass at P$(\vec{r}')$, contains a diverging kernel that is difficult to manage. Except for the Cartesian geometry \citep{macmillan1930theory,wal76}, there is no closed-form for the above integral, and one must therefore employ specific techniques, which is rarely done \cite[e.g.][]{ansorg03,hurepierens05,lietal09}. The singularity is usually avoided by raising the separation as follows \citep[e.g.][]{binneytremaine87,paplin89,pa12}
\begin{equation}
|\vec{r}-\vec{r'}| \leftarrow \sqrt{|\vec{r}-\vec{r'}|^2+\lambda^2} \ge |\lambda| > 0,
\label{eq:lambda}
\end{equation}
where $\lambda$ is a ``softening'' length. A small value is usually considered, typically the numerical resolution or the smallest scale available. This bare substitution does not economise the numerical quadrature, while replacing Eq.(\ref{eq:psicell}) by an analytical function offers much more flexibility. A common choice is the ``softened'' point mass potential
\begin{equation}
-\frac{Gm}{\sqrt{|\vec{r}-\vec{r_0'}|^2+\lambda^2}} \equiv \psiplum(\vec{r};\vec{r_0'},\lambda),
\label{eq:psiplummer}
\end{equation}
also known as the Plummer potential for a sphere with centre $\vec{r_0'}$, mass $m$, and core radius $\lambda$ \cite[e.g.][]{dejonghe87}. In disc simulations, the role of the parameter is to account for the vertical extension of matter, and $\lambda$ is thereby set to some fraction of the local thickness. In practical, authors often choose a constant value of the order of $0.6$, to a factor $2$ typically \citep[][]{masset02,bmp11,merubate12,mukm12}, and sometimes a function of space \citep[see e.g. Tab. 1 in][]{hurepierens05}.

It is clear that Eq.(\ref{eq:psiplummer}) does not corresponds precisely to the Newtonian potential of a numerical cell, unless $\lambda$ satisfies
\begin{equation}
\psiplum(\vec{r};\lambda) - \psicell(\vec{r}) =0,
\label{eq:rootfinding}
\end{equation}
everywhere, i.e.
\begin{equation}
\lambda^2 = \left(\frac{Gm}{\psicell}\right)^2 - |\vec{r}-\vec{r_0'}|^2,
\label{eq:root}
\end{equation}
which is inevitably a complicated function of space and cell's parameters (geometry/shape and density). This is especially true inside the cell and in close neighbourhood. At long range, the requested $1/|\vec{r}-\vec{r_0'}|$ decline typical of Newtonian gravity is however achieved for any small value of the softening length as an expansion of Eq.(\ref{eq:psiplummer}) in $\lambda^2/|\vec{r}-\vec{r_0'}|^2 \ll 1$ shows. At first sight, it could seem superfluous to try to ameliorate this parameter, but there is an essential point: in a discretized disc, almost all the grid cells are located far away from a given node P$(\vec{r})$, meaning that most cells are seen as distant cells. Finally, it is rather risky to be satisfied with a small but arbitrary value, in particular if a certain level accuracy is requested.

In this paper, we show that, if $\lambda$ is defined to faithfully reproduce the Newtonian potential of a cell, then not only it depends on the cell's parameters and distance, but it can be an imaginary number, which is not common at all. A calculus aiming to extract the dipolar term in Eq.(\ref{eq:psicell}) | $\lambda$ in fact measures how $\psicell$ deviates from the monopole | is currently under way and will be the aim of an dedicated article. In the meanwhile, we present a simple method to estimate $\lambda$, combining some analytics and a simple numerical scheme. We work in cylindrical geometry, well suited to rotating systems like discs.

The article is organized as follows. In Section \ref{sec:ir}, we show in one dimension that the softening length can be either a real or imaginary number. This result is supported by a analytical development valid at long-range. In Section \ref{sec:cylindricalcells}, we illustrate the complicated topology of $\lambda$ associated with the potential of the cylindrical cell. By using Simpson's rule, we determine the softening length with good accuracy. We then propose a new prescription valid for almost any distance and cell's geometry (rigorously, with radial-to-azimuthal aspect ratio around unity). We show in Sect. \ref{sec:newp} through a few concrete tests (regular and log. grid spacings, flared power-law discs, fully inhomogeneous density) that this new prescription gives much better results than the standard prescription. A concluding section summarizes the results. A few extensions of this work are discussed.

\begin{figure}
\includegraphics[width=11cm,bb=0 0 466 369,keepaspectratio=true]{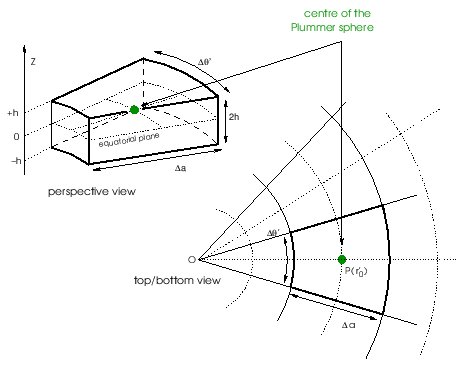}
\caption{Cylindrical cell (3D-view and projection in the midplane), centre of the Plummer sphere and associated notations.}
\label{fig:cells.eps}
\end{figure}

\section{Evidence for imaginary softening lengths}
\label{sec:ir}

In order to estimate $\lambda$ from Eq.(\ref{eq:root}), we work in cylindrical coordinates well suited to rotating discs. The system is supposed to be discretized on a numerical grid where each cell is a cylindrical sector, as the one depicted in Fig. \ref{fig:cells.eps}. The cell has centre $\vec{r}'_0(a_0,\theta'_0,z_0)$, angular extension $\Delta \theta'$, radial size $\Delta a$ and height $2h$. At this level, there is no particular constraint on the shape paremeters (i.e. vertical-to-radial and radial-to-azimuthal extension ratios). We however make the assumption that the fluid density does not vary inside. The elementary mass is $dm'=\rho a da d\theta' dz$, and $\rho$ is the constant density. If we set
\begin{equation}
\icell= \int_{z_0-h}^{z_0+h}dz\underbrace{\int_{a_0-\frac{1}{2}\Delta a}^{a_0+\frac{1}{2}\Delta a}da\underbrace{\int_{\theta'_0-\frac{1}{2}\Delta \theta'}^{\theta'_0+\frac{1}{2}\Delta \theta'}{\frac{ad\theta'}{|\vec{r}-\vec{r}'|}}}_{\iarc}}_{\ipolar},
\label{eq:icell}
\end{equation}
then $\psicell$ is just
\begin{equation}
\psicell=-\frac{Gm}{V}\icell,
\end{equation}
where $m=\rho V$ is the mass of the cell and $V$ its volume. Then, according to Eq.(\ref{eq:root}), the softening length is
\begin{equation}
\lambda^2 = \left(\frac{V}{\icell}\right)^2 - |\vec{r}-\vec{r_0'}|^2.
\label{eq:root_cell}
\end{equation}
To anticipate, we see that $\lambda^2$ can be negative, which implies an imaginary softening length. This situation means that the point mass potential is weaker than that of the cell, and must therefore be increased (by decreasing the separation $|\vec{r}-\vec{r}'|$).

\subsection{Exact formula for the massive arc}

We first integrate the kernel over the azimuth, i.e. calculate $\iarc$ in Eq.(\ref{eq:icell}). This quantity is linked to the Newtonian potential of a homogeneous, massive arc with radius $a$, length $\ell=a \Delta \theta'$, and opening angle $\Delta \theta'$, namely
\begin{equation}
\psiarc(\vec{r})=-\frac{Gm}{\ell}\iarc.
\end{equation}
It is traditionally expressed in terms of incomplete elliptic integral of the first kind
\begin{equation}
F(k,\phi) = \int_0^\phi{\frac{d\theta'}{\sqrt{1-k^2 \sin ^2 \theta'}}},
\end{equation}
where $k \le 1$ is the modulus. With the notations P$(\vec{r})\equiv(R, \theta, Z)$, P$'(\vec{r}')\equiv(a, \theta', z)$, we have
\begin{equation}
|\vec{r}-\vec{r}'|^2 = (a+R)^2+\zeta^2-4aR \sin^2 \phi,
\end{equation}
where $\zeta \equiv Z-z$ and $2\phi=\pi+\theta-\theta'$, and so the modulus is given by
\begin{equation}
k^2[(a+R)^2+\zeta^2] = 4aR. 
\label{eq:k}
\end{equation}
We then have \citep{pz05,thh2014Cemda}
\begin{equation}
\iarc = \sqrt{\frac{a}{R}}k \Delta F,
\label{eq:iarc2}
\end{equation}
where $\Delta F(k;\phi_1,\phi_2) \equiv F(k,\phi_1)- F(k,\phi_2)$, $2\phi_1=\pi+\theta-\theta'_0+\frac{\Delta \theta'}{2}$, and $2\phi_2=\pi+\theta-\theta'_0-\frac{\Delta \theta'}{2}$. So, the softening length associated with the arc is given by
\begin{equation}
\lambda^2 = \left(\frac{\ell}{\iarc}\right)^2 - |\vec{r}-\vec{r_0'}|^2.
\label{eq:root_arc}
\end{equation}
 By using Eq.(\ref{eq:iarc2}), we find for $R>0$
\begin{eqnarray}
\label{eq:lambdaforanarc}
\lefteqn{\frac{\lambda^2}{4aR} =\frac{1}{k^2}\left[\left(\frac{\Delta \theta'}{2\Delta F}\right)^2-1\right]+\sin^2\phi_0,}\\
\nonumber
&&\equiv \bar{\lambda}^2
\end{eqnarray}
where $2\phi_0=\pi+\theta-\theta_0'$. The interpretation of this result is the following: {\it the Plummer potential computed with $\lambda$ given by Eq.(\ref{eq:lambdaforanarc}) perfectly  reproduces the Newtonian potential of a homogeneous arc everywhere in space}. The occurrence of imaginary numbers for $\lambda$, already perceptible in Eq.(\ref{eq:root_cell}), can be understood from this latter formula: we have $F(k,\phi) \ge \phi$ whatever $k$ and $k \le 1$, which means that weight is possibly given to the negative term $-1/k^2$.

\begin{figure}
\includegraphics[width=11cm,bb=0 0 460 457,keepaspectratio=true]{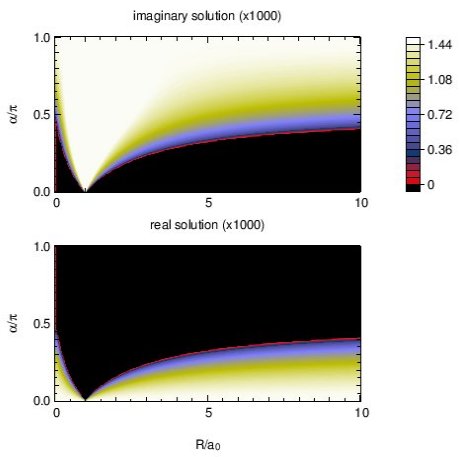}
\caption{Value of $\bar{\lambda}$ computed in the equatorial plane $\zeta_0=0$ from Eq.(\ref{eq:lambdaforanarc}): imaginary solutions ({\it top}) and real solutions ({\it bottom}). The arc (not visible) has coordinates $R/a_0=1$, $z=z_0$ and $\alpha=0$, and opening angle $\Delta \theta'=0.01$. Only the half-plane $\alpha \in [0,\pi]$ is shown (because of symmetry). According to Eq.(\ref{eq:lambdaapprox1}), we have here $\bar{\lambda} \approx 0.144 \times \Delta \theta'\sqrt{\cos \alpha}$ at large separation $R/a_0$.}
\label{fig:arc.eps}
\end{figure}

Figure \ref{fig:arc.eps} shows the normalized softening length  $\bar{\lambda}$ versus $R/a_0$ and $\theta - \theta_0' \equiv \alpha$ (in units of $\pi$) for an arc with opening angle $\Delta \theta'=0.01$ (this value corresponds to a resolution number $N_\theta \equiv 2\pi/\Delta \theta' \approx 629$ which is typical in high resolution disc simulations). Depending on the field point, the softening length is either a real or a pure imaginary number. We see that it is also a complicated function of space, especially at short radii and in the vicinity of the arc (see the pinching in the figure at $R/a_0=1$ and $\alpha \equiv \theta-\theta'_0=0$). In contrast, for $R/a_0 \gg 1$ (i.e. at long range), $\bar{\lambda}$ is mainly a function of $\alpha $. At long-range, imaginary numbers are found for $|\alpha| \ge \frac{\pi}{2}$ typically, which corresponds to field points located in the concave side of the arc (while real values are for the convex side of the arc). The length is zero for $\alpha= \pm \frac{\pi}{2}$, which corresponds to the direction of the arc: matter located beyond and below the centre $\vec{r_0'}$ compensate somehow, and do not contribute at the actual expansion order.

\subsection{Long-range behavior for the massive arc}
\label{subsec:lrarc}

We can precisely catch the long-range behavior of $\lambda$  by writing $F(k,\phi)$ as a double series in $k$ and $\phi$ around $k \rightarrow 0$, but an expansion up to order $3$ is necessary here \citep{gradryz07}. One goes more straight to the point by expanding the kernel $1/|\vec{r}-\vec{r'}|$ over $\alpha = \theta-\theta'_0$ before integration in azimuth. After some algebra detailed in the Appendix \ref{sec:app1}, we find
\begin{equation}
\label{eq:i1approx}
\iarc = \frac{a_0\Delta \theta'}{|\vec{r}-\vec{r_0'}|} \left[1 - \frac{aR}{|\vec{r}-\vec{r_0'}|^2} \cos \alpha \left(1 - \sinc \frac{\Delta \theta'}{2} \right) \right].
\end{equation}
 For $\Delta \theta'/2 \ll 1$, which corresponds to arcs with a small opening angle, an expansion of the sine cardinal function leads to
\begin{equation}
\bar{\lambda}^2 = \frac{1}{2} \left(1 - \sinc{\frac{\Delta \theta'}{2}}\right)\cos \alpha  \approx \frac{\Delta \theta'^2}{48}\cos \alpha,
\label{eq:lambdaapprox1}
\end{equation}
This is in agreement with what is observed in Fig. \ref{fig:arc.eps} for $R/a_0\gg 1$: at long-range, the length is independent on the radius and varies as $\sqrt{\cos(\theta-\theta'_0)}$ where the cosine can be positive or negative. 

\section{The case of cylindrical cells}
\label{sec:cylindricalcells}
 
\subsection{A numerical example}

\begin{figure}
\includegraphics[width=11cm,bb=0 0 476 456,keepaspectratio=true]{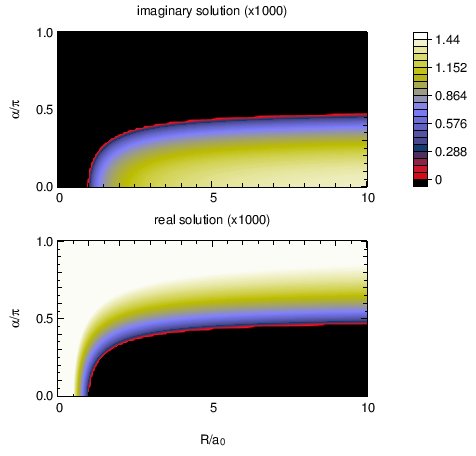}
\caption{Same legend and same color code as for Fig. \ref{fig:arc.eps}, but for the cylindrical cell with opening angle is $\Delta \theta'=0.01$, radial extension $\Delta a=0.01$ and total height $2h=0.01$. The coordinates of the cell's centre are $R/a_0=1$, $\alpha=0$ and $z_0$. Zones where $\lambda$ is real/complex are inverted with respect to the $1$D case (see Fig. \ref{fig:arc.eps}; see text).}
\label{fig:3dcell.eps}
\end{figure}

\begin{figure}
\includegraphics[width=11cm,bb=0 0 475 460,keepaspectratio=true]{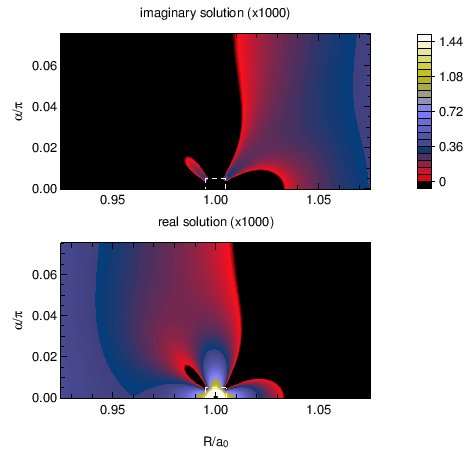}
\caption{Same legend and same color code as for Fig. \ref{fig:3dcell.eps}, but zoomed around the numerical cell (boundary in dashed white line). Interior values are out of range of the color code ($\bar{\lambda} \sim 0.0021$ at the centre).}
\label{fig:3dcell_zoomed.eps}
\end{figure}

 A closed-form for the potential of a homogeneous cylindrical cell, i.e. for $\icell$, is apparently missing yet so there is no equivalent to Eq.(\ref{eq:lambdaforanarc}). We can employ numerical means\footnote{In practical, the potential $\psicell$ is determined from the contour integral reported by \cite{thh2014Cemda}. In the paper throughout, we use this accurate method to generate reference values. \label{note:refpot}}. An illustration is given at Fig. \ref{fig:3dcell.eps} for a cylindrical cell with parameters $a_0  \Delta \theta'=\Delta a=2h=0.01$. We see that zones where the length is real and imaginary are still present, while inverted with respect to the arc. The softening length is now imaginary at large radii  for $|\alpha| \la \frac{\pi}{2}$. Again, the map is more complex in the vicinity of the cell, where real values are found inside the cercle with radius $R \le a_0$ typically. Figure \ref{fig:3dcell_zoomed.eps} is a zoom around $\vec{r}_0$ and we see that imaginary numbers are indeed located at $R/a_0 \ga 1$. We perceive dipole and quadrupole features which are very difficult to catch by analytical means\footnote{There are two possible ways to get $\lambda$ inside the cell and neighbourhood: from the formula for the potential of the polar cell \citep{hure12}, or from the double expansion of $F(k,\phi)$ around $k=1$ and $\phi=\frac{\pi}{2}$ \citep{vdv69}. Despite efforts, we failed in producing a result. This question remains open. \label{note1}}. Inside the cell, $\lambda$ is real and close to spherical symmetry\footnote{The potential of a perfect cube is very close to that of a sphere, and there is no dipole contribution \citep{durand64}.}. Just beyond the corners of the cell, the magnitude of the point mass potential is weaker, and must therefore be raised via an imaginary length, which explains the four lobes. By no means, $\lambda \propto h$ here.

The softening length depends on several factors linked to the cell's geometry, namely
\begin{itemize}
\item the polar shape factor $\frac{a_0\Delta \theta'}{\Delta a}$,
\item the meridional shape factor $\frac{2h}{\Delta a}$,
\end{itemize}
and on the relative altitude $Z/h$ too. Exploring the solution of Eq.(\ref{eq:rootfinding}) by varying these quantities would be interesting but it is out of the scope of this paper. It is however worth noting that $\lambda$ is particularly sensitive to $\frac{2h}{\Delta a}$ when this ratio is around unity. Actually, as the meridional shape factor is larger than unity and decreases, the solution has first a spherical geometry and $\lambda$ is real. When $\frac{2h}{\Delta a}$ reaches unity (as in the canonical example above), a quadrupolar pattern quickly sets in (see Fig. \ref{fig:3dcell_zoomed.eps}), while imaginary solutions appear at $R/a_0 > 1$. Below unity, $\lambda$ is mainly an imaginary number, except in the very close vicinity of the cell (and inside it) where $\lambda$ remains real. The real-to-imaginary swing occurs for a variation of $10\%$ of $\frac{2h}{\Delta a}$ around unity typically.

\subsection{Central values}

\begin{figure}
\includegraphics[width=11cm,bb=0 0 503 364,keepaspectratio=true]{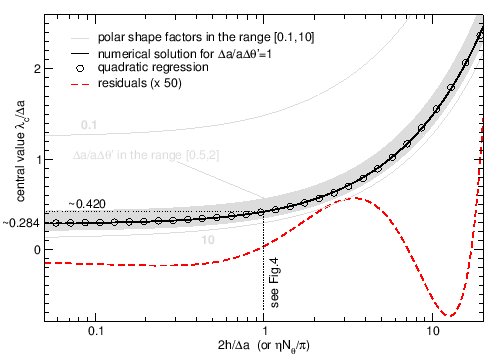}
\caption{Central value of the softening length versus the meridional shape factor $2h/\Delta a$ of the grid, in units of the radial extension $\Delta a$ (which is held fixed), for $\frac{a_0\Delta \theta'}{\Delta a}=1$ ({\it black}). The quadratic regression (with associated residuals) refers to Eq.(\ref{eq:lcfitted}). Also indicated ({\it grey}) is the central value when varying the polar shape factor $\frac{a_0\Delta \theta'}{\Delta a}$ in the range $[0.1,10]$; see Eq.(\ref{eq:lcfitted_gen}) for the double fit.}
\label{fig:2hoverda.eps}
\end{figure}

The value $\lambda_c \equiv \lambda(\vec{r}=\vec{r}_0)$ computed at the cell's centre is plotted versus the meridional shape factor $\frac{2h}{\Delta a}$ in Fig. \ref{fig:2hoverda.eps}, while holding $\frac{a_0\Delta \theta'}{\Delta a}=1$ (which is a common set up in disc simulations). A parabolic fit, valid for $\frac{2h}{\Delta a} \in [0,30]$, gives
\begin{equation}
\label{eq:lcfitted}
\frac{\lambda_c}{\Delta a} \approx 0.28622 + 0.13457 \frac{2h}{\Delta a} - 0.0013549 \left(\frac{2h}{\Delta a}\right)^2,
\end{equation}
with an error of the order of a few percents. The dependency on $\frac{a_0\Delta \theta'}{\Delta a}$, also indicated in the figure, is weak within reasonnable limits (see the Appendix \ref{sec:app2} for a double fit of $\lambda_c$). Note that in the case where $h \propto \Delta a$, then $\lambda_c \propto h$ provided $\frac{2h}{\Delta a} \ll 1$. This special case concerns cylindrical grids with logarithmic spacing in radius and cells having a height linearly increasing with radius. However, we see that $\lambda_c$ is generally not proportional to the cell's height $2h$, and this is particularly true when $\frac{a_0\Delta \theta'}{\Delta a} \ne 1$. This result also goes against the standard prescription stating that $\lambda \propto h$.

\subsection{Reliable estimate from Simpson's rule}

\begin{figure}
\includegraphics[width=11cm,bb=0 0 477 252,keepaspectratio=true]{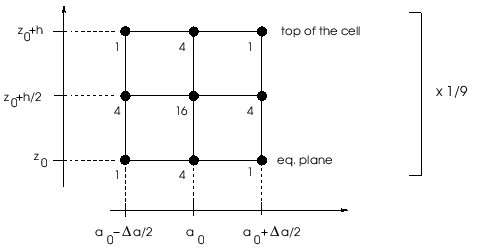}
\caption{Stencil and $w_{ij}$for Simpson's rule in two dimensions (see also Fig. \ref{fig:cells.eps}).}
\label{fig:stencil.eps}
\end{figure}

In the midplane of the cell, i.e. at $Z=z_0$, we can rewrite Eq.(\ref{eq:icell}) as follows
\begin{equation}
\icell= 2 \int_{z_0}^{z_0+h}dz\int_{a_0-\frac{1}{2}\Delta a}^{a_0+\frac{1}{2}\Delta a}{\iarc(a,z)da}.
\label{eq:icell_sym}
\end{equation}
where $\iarc$ must be regarded as function of $a$ and $z$ (it also depends on $\theta_0$ and $\Delta \theta'$ and $\vec{r}$ as well). Because $\iarc$ is essentially regular\footnote{In practical, $\iarc$ has large amplitude in the very close vicinity of the arc, and not always monotonic. Even, it is logarithmically diverging as soon as the modulus $k=1$, which occurs for $R=a_0$ for $Z=z_0$. To avoid singular values, we can safely replace $k$ by $k_*$ in Eq.(\ref{eq:k}), where
\begin{equation}
k_*^2[(a_0+R)^2+\zeta_0^2+\epsilon^2] = 4a_0R,
\label{eq:kstar}
\end{equation}
and $\epsilon = \frac{1}{100}\min\{\ell_1,\ell_2,\ell_3\}$ for instance. \label{note:kstar}}, we can employ a classical quadrature scheme to estimate $\icell$, i.e. the double integral is replaced by a discrete summation with values of $\iarc$ taken at different places $(a,z)$. For the present purpose, Simpson's rule has a sufficient accuracy. There are therefore $9$ nodes $(a_i,z_j)$ involved in total, namely
\begin{itemize}
\item $a_i=a_0+(i-2)\frac{\Delta a}{2}$ with $i=1,2,3$,
\item $z_j=z_0+(j-1)\frac{h}{2}$, with $j=1,2,3$.
\end{itemize}
We then have
\begin{equation}
\icell \approx h \Delta a \times \frac{1}{18}\sum_{i=1}^9{w_{ij} \iarc(a_i,z_j)},
\label{eq:icell_simpson}
\end{equation}
where the corresponding weights $w_{ij} \in \{1,4,16\}$ are given by the stencil in Fig. \ref {fig:stencil.eps}. The softening length is then deduced from Eqs.(\ref{eq:root_cell}) and (\ref{eq:icell_simpson}) where $V=a_0 \Delta \theta' h \Delta a $. According to Simpson's rule, the error when estimating $\icell$ from Eq.(\ref{eq:icell_simpson}) is of the order of $(h \Delta a)^5$. This makes this approach reliable enough and relatively general, at least for numerical applications based on fine cylindrical grids.

\begin{figure}
\includegraphics[width=11cm,bb=0 0 462 464,keepaspectratio=true]{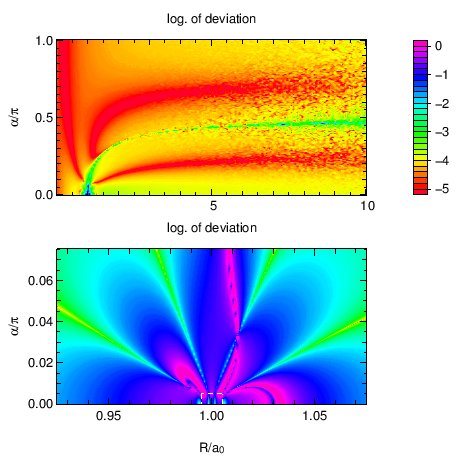}
\caption{Relative error (decimal log. scale) when computing the softening length $\lambda$ from Eq.(\ref{eq:icell_simpson}) at large scale ({\it top panel}), and in the vicinity of the cell at $\vec{r}_0$. The conditions are the same as for Figs.  \ref{fig:3dcell.eps} and \ref{fig:3dcell_zoomed.eps}.}
\label{fig:eindex.ps}
\end{figure}

We have computed  $\lambda$ from Eq.(\ref{eq:icell_simpson}) and compared it with the reference$^{\ref{note:refpot}}$ for the same cell's parameter as considered in the previous example. The decimal log. of the relative error is displayed in Fig. \ref{fig:eindex.ps}. We see that accuracy is better than $10^{-4}$, except along the line where $\lambda=0$ and especially in the vicinity of the cell itself where this is rather $10\%$.

\section{What does this change for disc simulations ?}
\label{sec:newp}

\subsection{A new prescription}

On the basis of the results above and various tests performed, we can propose a new prescription for $\lambda$ which makes the softened point mass potential very close to the Newtonian potential of a cylindrical cell by combining: i) the numerical value computed at the cell's centre (the fit has sufficient accuracy), and ii) the $9$-point stencil. For a given cell with shape parameters $(\ell_1,\ell_2,\ell_3)$ and centre $\vec{r}_0$, the recipe is as follows:
\begin{itemize}
\item  $\lambda$ is given by Eq.(\ref{eq:lcfitted}) or Eq.(\ref{eq:lcfitted_gen}), if $\vec{r} = \vec{r}_0$,
\item  otherwise, $\lambda$ is given$^{\ref{note:kstar}}$ by Eqs.(\ref{eq:root_cell}) and (\ref{eq:icell_simpson}).
\end{itemize}

This composite formula is a priori easy to include into numerical codes.

\subsection{Two numerical tests}

By adding the contribution of individual cells, one can determine the total potential of any kind of inhomogeneous system, where each cell has its own shape parameters, density and associated softening length. We have therefore performed several concrete tests, in order to check the efficiency of this new prescription for $\lambda$. It is in particular interesting to compare with the standard prescription where $\lambda\propto h$.

Among possibilities, we have selected two configurations: i) a fully homogeneous and flat disc, and ii) a power-law density disc with constant aspect ratio. In both cases, the disc is axially symmetrical, and has inner edge at $\ain=0.5$. The numerical grid consists of $N_R$ nodes in radius, and $N_\theta$ node in azimuth. In the vertical direction, there is a unique node located in the disc midplane. There are therefore $(N_R-1) (N_\theta-1)$ cells in total and the same amount of nodes. In practical, we use $N_\theta=2 \times N_R = 64$ (which leads to $\Delta \theta' \approx 0.1$; see below for higher resolutions). In addition, we have considered two types of grid: one grid with regular spacing, and one with log. spacing. In this latter case, we have  the relationship
\begin{equation}
\frac{2h}{\Delta a} = \frac{\eta}{\pi} N_\theta,
\end{equation}
where  $\eta = h/a$ is the disc aspect ratio.

\begin{figure}
\includegraphics[width=11cm,bb=0 0 507 428,keepaspectratio=true]{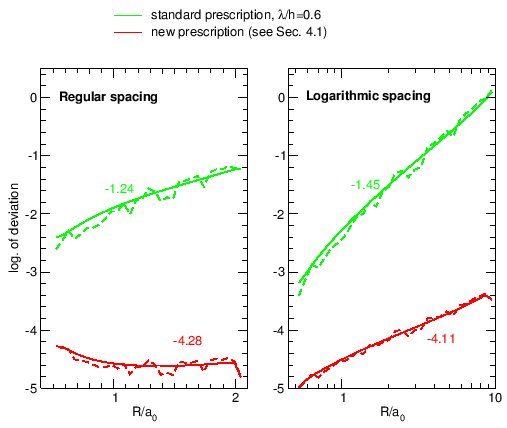}
\caption{Relative error for potential values (decimal log. scale) with respect to the reference (see text) for the standard prescription with $\lambda/h=0.6$ ({\it green}) and for the present prescription (see Sec. \ref{sec:cylindricalcells}.1). The disc is fully homogeneous and flat with inner edge at $\ain=0.5$ ({\it plain lines}); the azimuthal number is $N_\theta=64$ and radial number $N_R=N_\theta/2$. Two grids are used: regular spacing ({\it left panel}) and log. spacing ({\it right panel}). The label on the curves give the average deviation. For the fully inhomogeneous density field ({\it dashed lines}; see text), errors are azimuthally averaged.}
\label{fig:changes_random.eps}
\end{figure}

At each radius $R$ in the midplane, the reference potential is reconstructed by adding individual contributions$^{\ref{note:refpot}}$. In parallel, we perform the same summation but using the Plummer potential i) with the new prescription, and ii) with the standard prescription where $\lambda \propto h$ (in practical, we take $\lambda=0.6h$). Figure \ref{fig:changes_random.eps} shows the deviations observed for the homogeneous flat disc. We see that this an improvement by $2-4$ digits depending on the radius in the disc. Results are much better close to the outer edge. We have also considered an inhomogeneous verision by simply multiplying the density field of each cell by a random factor in the range $[1,2]$. The results are plotted in the same figure and the conclusions are unchanged.

\begin{figure}
\includegraphics[width=11cm,bb=0 0 500 431,keepaspectratio=true]{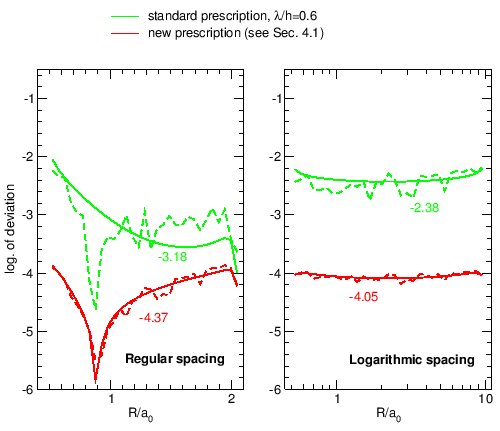}
\caption{Same legend and same conditions as for Fig. \ref{fig:changes_random.eps} but for the flared, power-law density disc (see text).}
\label{fig:changes_plf.eps}
\end{figure}

Figure \ref{fig:changes_plf.eps} shows the same results but for the flared, power-law density disc where $\frac{h(a)}{a}=\frac{1}{2}\Delta \theta'$ and $\rho(a) \propto a^{-2.5}$, which is typical. The new prescription is again better, while the benefit is a little bit reduced for the grid with regular spacing. For the grid with log. spacing, potential values are improved by almost $2$ more digits.

\subsection{Resolution effects}

\begin{figure}
\includegraphics[width=11cm,bb=0 0 512 388,keepaspectratio=true]{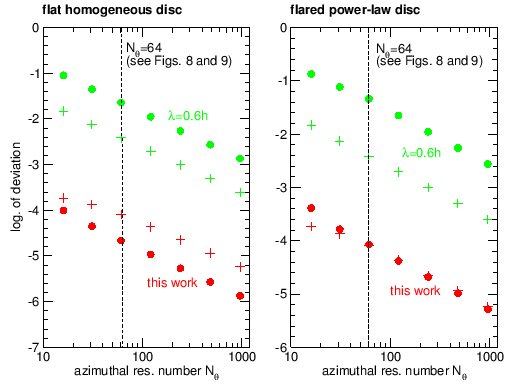}
\caption{Same legend as for Fig. \ref{fig:changes_random.eps} and \ref{fig:changes_plf.eps} but for various azimuthal resolution numbers $N_\theta$. The deviations are computed at the radius located at half-sampling, using the two kinds of grid: regular spacing ({\it plain circles}) and log. spacing ({\it crosses}).}
\label{fig:changes_vares.eps}
\end{figure}

Figure \ref{fig:changes_vares.eps} gives the deviation observed at half-sampling in radius for $N_\theta \in \{32,64,128,256,512,1024\}$ while maintaining $N_R=N_\theta/2$ and $\ain=0.5$, for the two configurations considered above. This confirms that accuracy can be significantly improved by $2-3$ orders of magnitude with respect to the standard case where $\lambda \propto h$ whatever the resolution.

\section{Concluding remarks}

In this article, we have proposed a new prescription for the softening length which enables to reproduce the Newtonian potential and the associated acceleration field from the Plummer formula with a very good level of accuracy. A major result is {\it the occurence of purely imaginary lengths}, which was not envisaged before. Not only the location in space but also the grid shape plays a decisive role. Besides, the nominal value is generally not a fraction of the local thickness (unless a very special tuning of the grid parameters), in constrast with the current standard.

This study is preliminary and can be continued in several ways. As annonced, a full dipolar expansion of the potential of the cylindrical cell is currently in progress. An analytical formula for $\lambda$ would actually be of great help not only to understand how the length depends on the cell's shape, but also to avoid the evaluation of the elliptic integrals $F(k,\phi)$. Actually, the extra computational cost when using Eq.(\ref{eq:icell_simpson}) and (\ref{eq:root_cell}) instead of a constant $\lambda/h$ is due to the evaluation of $F(k,\phi)$ ($18$ evaluations in total for each cell). If the grid is fixed, this can be done once for all, but it is is obvious that any approximation for this special function would reduce the computing time. It would also be interesting to understand how this precription is modified if vertical stratification of matter is accounted for (for instance with a Gaussian profile). Another challenge concerns the derivation of the softening length in the case of spherical and Cartesian geometries. In the meanwhile, we plan to investigate how this prescription impacts on hydrodynamical simulations since acceleration fields are much accurate with this method (Trova, Pierens, \& Hur\'e, 2014, in preparation).

\section*{Acknowledgments}

It is a pleasure to thank W. Kley, F. Hersant, and especially A. Pierens for valuable comments and advices. A. Trova is greateful to V. Karas, the Relativistic Astrophysics Group and the Academy of Sciences  in Prague. J.M. Hur\'e is greateful to the CNRS-INSU for support. We thank the anonymous referee for detailed reports and criticisms.

\bibliographystyle{mn2e}

\appendix


\section{Softening length associated with the massive arc. Long-range behavior}
\label{sec:app1}

The relative distance is given by
\begin{eqnarray}
\lefteqn{|\vec{r}-\vec{r'}|^2 = (R \cos \theta - a\cos \theta')^2}\\
\nonumber
&& \qquad  \qquad +(R \sin \theta - a \sin \theta')^2 + (Z-z)^2,\\
\nonumber
&& = (a+R)^2+\zeta^2-4aR \sin^2 \phi,
\end{eqnarray}
where $\zeta \equiv Z-z$ and $2\phi=\pi+\theta-\theta'$. If we introduce the angle $\alpha'=\theta'-\theta'_0$, then
\begin{equation}
2\phi_0=\pi+\theta-\theta'_0=\pi+\theta-\theta'+(\theta'-\theta'_0)=2\phi+\alpha',
\end{equation}
and so $\sin^2 \phi = \sin^2 \left(\phi_0-\frac{\alpha'}{2}\right)$. From simple trigometric rules, we find
\begin{eqnarray}
\lefteqn{|\vec{r}-\vec{r'}|^2  
 = D_1^2 - 4aR \sin\frac{\alpha'}{2} \sin \left(\frac{\alpha'}{2}-2\phi_0\right)},
\end{eqnarray}
where $D_1^2=(a+R)^2+\zeta^2-4aR \sin^2 \phi_0$. We can directly deduce $\lambda$ from Eq.(\ref{eq:lambda}). Provided $\frac{\alpha'}{2}\ll 1$, we have
\begin{equation}
\frac{1}{|\vec{r}-\vec{r'}|} \approx \frac{1}{D_1} \left[1 + \frac{2aR}{D_1^2}  \sin\frac{\alpha'}{2} \sin \left(\frac{\alpha'}{2}-2\phi_0\right) \right]
\end{equation}
Since
\begin{eqnarray}
\int{\sin\frac{x}{2} \sin \left(\frac{x}{2}-x_0\right)dx}=\frac{1}{2}\left[x \cos x_0 - \sin(x-x_0)\right],
\end{eqnarray}
and $d\theta'=d\alpha'$, the integral over the shifted azimuth $\alpha'$ has bounds $\pm \frac{\Delta \theta'}{2}$, and we have
\begin{eqnarray}
\nonumber
\lefteqn{\int_{-\frac{\Delta \theta'}{2}}^{+\frac{\Delta \theta'}{2}}{\sin\frac{\alpha'}{2} \sin \left(\frac{\alpha'}{2}-2\phi_0\right)d\alpha'}=\frac{\Delta \theta'}{2} \cos 2 \phi_0} \\
& \qquad\qquad\qquad\qquad\qquad\qquad \times \left(1 - \sinc \frac{\Delta \theta'}{2} \right),
\end{eqnarray}
If we now define the angle $\alpha=\theta-\theta'_0$ (see Sect. \ref{sec:ir}), $\cos 2\phi_0 = - \cos \alpha$, and so we have
\begin{eqnarray}
\label{eq:i1longrange}
\lefteqn{I_1 = \int_{\theta'_0-\frac{\Delta \theta'}{2}}^{\theta'_0+\frac{\Delta \theta'}{2}}{\frac{d\theta'}{|\vec{r}-\vec{r'}|}}}\\
&& \approx \frac{\Delta \theta'}{D_1} \left[1 - \frac{aR}{D_1^2} \cos \alpha \left(1 - \sinc \frac{\Delta \theta'}{2} \right) \right]
\end{eqnarray}
and so
\begin{eqnarray}
\lefteqn{\psiarc(\vec{r})= - G\frac{ma}{\ell} I_1}\\
&& \approx - \frac{Gm}{D_1} \left[1 - \frac{aR}{D_1^2} \cos \alpha \left(1 - \sinc \frac{\Delta \theta'}{2} \right) \right]
\end{eqnarray}
By setting $a=a_0$ and $z=z_0$ (so that the centre of the arc coincides with the centre of the Plummer sphere), we have $|\vec{r}-\vec{r_0'}|=D_1$, and we directly find from Eq.(\ref{eq:root})
\begin{equation}
\lambda^2 =   D_1^2 \left\{\frac{1}{\left[1 - \frac{aR}{D_1^2} \cos \alpha \left(1 - \sinc \frac{\Delta \theta'}{2} \right) \right]^2} - 1 \right\}
\end{equation}

By expanding the Plummer potential, we have
\begin{equation}
\psiplum(\vec{r};\vec{r_0'},\lambda) = -\frac{Gm}{|\vec{r}-\vec{r_0'}|}\left(1-\frac{\lambda^2}{2|\vec{r}-\vec{r_0'}|^2} + \dots \right),
\label{eq:psiplummer_approx}
\end{equation}
and so, by comparison, we have
\begin{equation}
\bar{\lambda}^2 =  \frac{1}{2}\cos \alpha \left(1 - \sinc \frac{\Delta \theta'}{2} \right) 
\end{equation}
where we have set $\bar{\lambda} \times 2 \sqrt{aR} \equiv \lambda$. For $\Delta \theta'/2 \ll 1$, which correspond to arcs with a small opening angle, we have
\begin{equation}
\bar{\lambda}^2 \approx \frac{1}{48}{\Delta \theta'}^2\cos (\theta-\theta'_0) 
\end{equation}
It turns out that $\lambda$ is real for $\theta-\theta'_0 \in [-\frac{\pi}{2},+\frac{\pi}{2}]$ and imaginary otherwise. It is zero for $\theta-\theta'_0=\pm \frac{\pi}{2}$.

\section{A fit for $\lambda_c$}
\label{sec:app2}

If we set $x=\frac{a_0\Delta \theta'}{\Delta a}$ and $y=\frac{2h}{\Delta a}$, a fit valid in the range $x \in [0.1,10]$ and $y \in [0,30]$ is
\begin{eqnarray}
\nonumber
\frac{\lambda_c}{\Delta a} \approx \frac{0.13418}{x} + 0.1402 + \left(\frac{0.0053327}{x} + 0.12983\right) y\\
 + \left(\frac{0.00012699}{x} - 0.0016083\right)y^2,
\label{eq:lcfitted_gen}
\end{eqnarray}

\end{document}